%% file: main.tex
\documentclass[conference]{IEEEtran}
\usepackage{amsmath,amsfonts}
\usepackage{algorithmic}
\usepackage{algorithm}
\usepackage{array}
\usepackage[caption=false]{subfig}
\usepackage{textcomp}
\usepackage{stfloats}
\usepackage{url}
\usepackage{verbatim}
\usepackage{graphicx}
\usepackage[nolist]{acronym}
\usepackage{cite}
\usepackage{svg}
\usepackage{soul}
\usepackage{color}
\usepackage{xcolor}
\usepackage[normalem]{ulem}
\usepackage{soulutf8}
\usepackage{orcidlink}

\input{acronyms}

\begin{document}

\title{
5G NR-V2X Scheduling Approaches for CPM Variable Size Traffic
}

\markboth{Journal of \LaTeX\ Class Files,~Vol.~14, No.~8, August~2021}
{Shell \MakeLowercase{\textit{et al.}}: A Sample Article Using IEEEtran.cls for IEEE Journals}

\author{
\IEEEauthorblockN{
Vittorio Todisco\IEEEauthorrefmark{1}\IEEEauthorrefmark{2}\orcidlink{0000-0002-6737-6710},
Mattia Andreani\IEEEauthorrefmark{3}\IEEEauthorrefmark{4}\orcidlink{0009-0008-8282-5343},
Maria Luisa Merani\IEEEauthorrefmark{3}\IEEEauthorrefmark{4}\orcidlink{0000-0003-4162-5042},
and Alessandro Bazzi\IEEEauthorrefmark{1}\IEEEauthorrefmark{2}\orcidlink{0000-0003-3500-1997}
}

\medskip
\IEEEauthorblockA{\IEEEauthorrefmark{1}Department of Electrical, Electronic and Information Engineering (DEI), University of Bologna, Italy}
\IEEEauthorblockA{\IEEEauthorrefmark{2}National Laboratory of Wireless Communications (WiLab), CNIT, Italy}
\IEEEauthorblockA{\IEEEauthorrefmark{3}Department of Engineering ``Enzo Ferrari'', University of Modena and Reggio Emilia, Italy}
\IEEEauthorblockA{\IEEEauthorrefmark{4}Consorzio Nazionale Interuniversitario per le Telecomunicazioni (CNIT), Italy}
}

\maketitle

\begin{abstract}
Collective perception messages (CPMs) introduce significant packet size variability due to dynamic object inclusion and periodic security overhead. While 5G NR-V2X Mode 2 typically employs semi-persistent scheduling (SPS) designed for periodic traffic with relatively stable packet sizes, the impact of realistic CPM-driven size fluctuations on distributed resource allocation remains insufficiently understood. This paper presents a comparative system-level evaluation of NR-V2X Mode 2 scheduling strategies under variable-size CPM traffic reconstructed from real-world perception datasets. We analyze dynamic scheduling and multiple SPS-based approaches, including padding-based allocation, aggressive size-driven reselection, and modulation and coding scheme (MCS) adaptation. Results show that packet size variability can significantly degrade reliability when scheduling stability is compromised. In particular, dynamic scheduling and aggressive reselection increase collision probability due to frequent resource reallocations. In contrast, SPS with padding converges to a stable resource allocation and provides robust performance, while MCS adaptation achieves the highest average packet reception ratio but with uneven reliability across packet types. The findings demonstrate that, under realistic CPM traffic, stability of resource usage is more critical than instantaneous load optimization, and provide design guidelines for CPS deployment over NR-V2X Mode 2.    
\end{abstract}

\bigskip

\section{Introduction}

\Ac{V2X} communication is a key enabler of cooperative and automated mobility, extending situational awareness beyond the limits of on-board sensors. \acp{C-ITS} are evolving from periodic status exchange toward services such as collective perception, where vehicles share information about detected objects in their surroundings. By disseminating object-level environmental data, the \ac{CPS} supports enhanced safety and automated driving functionalities that require awareness beyond line-of-sight \cite{bib:WolfCPS2015, bib:WolfCPS2016}.

\medskip
Compared to early periodic awareness messages, perception-driven traffic exhibits fundamentally different characteristics. The amount of information to be transmitted depends on the environment, traffic conditions, and sensor observations, leading to packet sizes that vary over time and across vehicles. These variations are structured rather than random: consecutive transmissions are often temporally correlated, reflecting the evolution of the perceived scene. In addition, security procedures such as periodic certificate transmissions introduce further, non-negligible size fluctuations that directly affect the transport block.

5G \ac{NR}-\ac{V2X} sidelink Mode~2 has been designed to support distributed \ac{V2X} communication without network assistance. In this mode, user equipment (UE) autonomously selects and reserves time--frequency resources based on channel monitoring and distributed resource exclusion procedures. \ac{SPS} is commonly adopted for periodic services, as it enables predictable resource reuse and reduces contention. However, \ac{SPS} implicitly assumes relatively stable packet sizes and transmission periodicity. When packet size fluctuates between consecutive transmissions, previously reserved resources may no longer match the required transport block size. This mismatch can trigger resource reselection or adaptation mechanisms, potentially increasing collision probability and degrading reliability.

Many existing studies simplify this problem in opposite directions. Some assume periodic traffic with constant or quasi-constant packet sizes, aligning well with the design assumptions of \ac{SPS} \cite{bib:Luca_fixedsize}. Others model fully aperiodic traffic with independent packet sizes, effectively considering worst-case variability \cite{bib:Luca_variablesize, bib:Baldo}. Real perception-driven traffic lies between these extremes: packet sizes evolve with temporal structure and correlation, yet still exhibit significant dynamics that may stress distributed resource allocation.

In this paper, we investigate how \ac{NR}-\ac{V2X} Mode~2 scheduling performs under realistic variable-size traffic conditions. Using \acp{CPM} \cite{bib:ETSI_Cpm} reconstructed from object traces recorded by a vehicle driving on a highway~\cite{bib:Cirrus_dataset}, we capture both temporally correlated payload variations and periodic security-induced size fluctuations. We conduct a comparative system-level evaluation of multiple scheduling strategies, including fully dynamic per-packet scheduling and different \ac{SPS}-based implementation choices such as padding-based allocation, aggressive size-driven reselection, and \ac{MCS} adaptation. We quantify their impact on reliability, resource stability, and channel utilization, and demonstrate that frequent resource reallocations triggered by packet size variability can significantly degrade performance.

Our findings highlight that, under realistic \ac{CPM} traffic, scheduling stability is more critical than instantaneous load optimization in distributed \ac{NR}-\ac{V2X} Mode~2 operation. These results provide practical guidance for configuring \ac{NR}-\ac{V2X} Mode~2 to support emerging cooperative perception services.


\section{Variable-Size CPM Traffic Model}

To evaluate scheduling strategies under realistic perception-driven traffic conditions, we reconstruct variable-size \acp{CPM} from object traces recorded by a vehicle driving on a highway \cite{bib:dataset}. The dataset captures the dynamic evolution of the surrounding environment, including the number and 
features
of detected objects over time \cite{bib:Cirrus_dataset}. Each trace sample contains object-level information such as position, velocity, and classification, which are encoded into \ac{CPM} payload elements according to the CPS message structure \cite{bib:ETSI_Cpm}.

The resulting packet size depends primarily on the number of objects included in each transmission. As traffic density and scene complexity vary, the number of reported objects fluctuates over time, leading to corresponding variations in payload size. These variations are temporally correlated: consecutive transmissions often contain similar object sets with incremental changes, reflecting the physical continuity of the driving scenario. Therefore, packet size evolution exhibits structured dynamics rather than independent random behavior.

In addition to object-dependent payload variability, security procedures introduce further size fluctuations. Periodic certificate transmissions add additional overhead to selected packets, increasing the required transport block size at regular intervals. This security-induced variability is superimposed on the object-driven payload evolution and may cause abrupt increases in packet size.

The overall packet size at transmission instant $k$ can therefore be expressed as
\begin{equation}
P_k = P_{\text{base}} + N_k P_{\text{obj}} + P_{\text{sec},k},
\end{equation}
where $P_{\text{base}}$ represents fixed header and management fields, $N_k$ is the number of objects included in the message, $P_{\text{obj}}$ denotes the average contribution per object, and $P_{\text{sec},k}$ models the periodic security overhead.

This modeling approach captures both temporally correlated payload dynamics and periodic size spikes due to certificate transmission. 
\begin{figure}[h!]
    \centering

\includegraphics[width=0.48\textwidth]{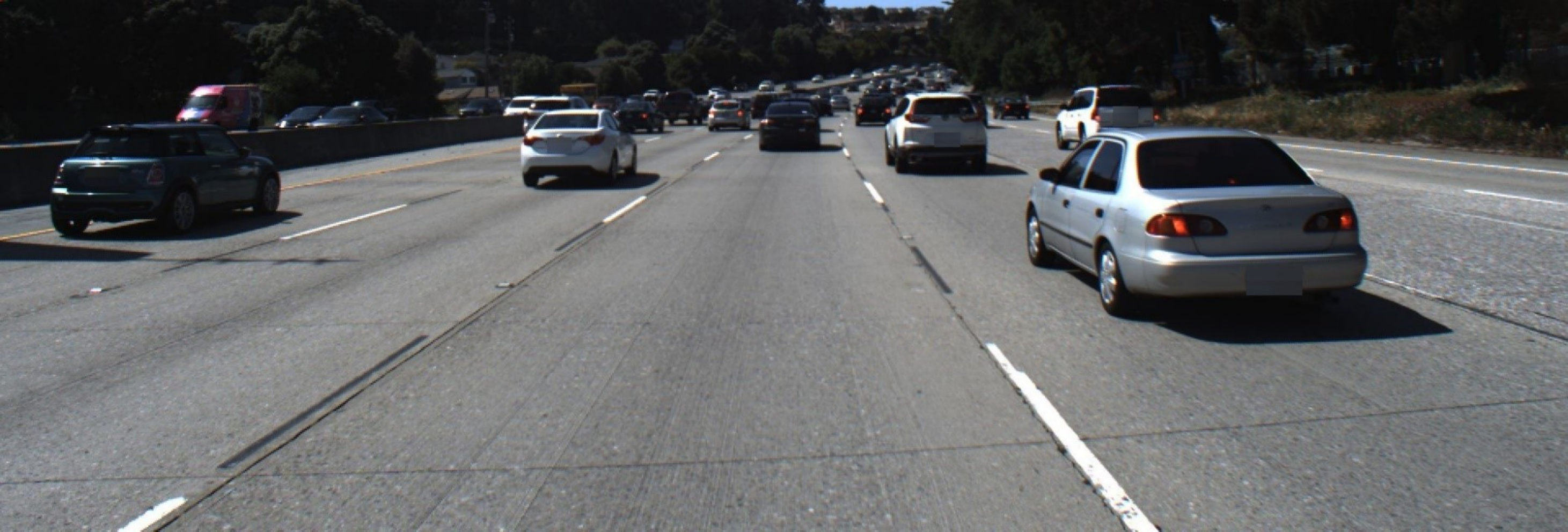}

    \caption{Camera frame from the scenario used for the study \cite{bib:Cirrus_dataset}.}
    \label{fig:cirrus_morecongested}
\end{figure}

\begin{figure}[t]
  \centering
  \includegraphics[width=0.88\columnwidth]{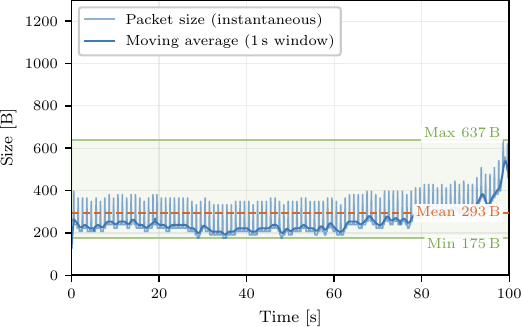}
    \vspace{-0.5em}
  \caption{Packet size over time during a non-congested driving period.}

  \label{fig:packet_size_100}
  
\end{figure}

\begin{figure}[t]
  \centering
  \includegraphics[width=0.88\columnwidth]{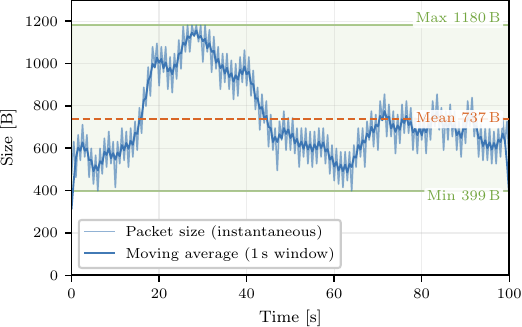}
  \vspace{-0.5em}
  \caption{Packet size over time during a congested driving period.}
  \label{fig:packet_size_500}
  \vspace{-1em}

\end{figure}

\subsection{Temporal Evolution of Reconstructed CPM Sizes}

Two representative portions of the highway dataset are selected. The first corresponds to a non-congested traffic condition, where the recording vehicle observes a limited number of surrounding vehicles. The second corresponds to a congested traffic condition, characterized by a significantly larger number of perceived objects.

For \ac{CPM} reconstruction, all detected objects in each frame are included in the generated message. No object filtering or relevance-based selection is applied. As a result, the reconstructed packet size reflects the full perception load observed in the dataset.
In the non-congested highway scenario, \acp{CPM} are generated with $T_{\mathrm{gen}} = 100$~ms, while in the congested scenario $T_{\mathrm{gen}} = 500$~ms is adopted to reduce overall channel load in the presence of a larger number of simultaneously transmitting vehicles.

Figs.~\ref{fig:packet_size_100} and~\ref{fig:packet_size_500} illustrate the temporal evolution of the reconstructed \ac{CPM} packet size for both driving period scenarios. Packet sizes are derived from the object count trace and mapped to \acp{CPM} according to the size model described above.
In our reconstruction, $P_{\text{obj}} = 16$~bytes per object, while $P_{\text{base}} = 53$~bytes aggregates fixed CPM and networking headers. The security overhead is modeled as $P_{\text{sec},k} \in \{231,\,89\}$~bytes, with a full certificate transmitted once per second and a digest included in intermediate messages.

For visualization and simulation input purposes, the 
reconstructed \ac{CPM} size is mapped to a valid \ac{NR} \ac{TB} size. This mapping assumes a reference \ac{MCS} equal to 12 and selects the smallest standardized \ac{TB} size, up to a maximum supported \ac{TB} size of $1180$ bytes under the considered configuration. This reference mapping is used only to obtain the effective Layer-2 packet size shown in the figures and does not constrain the scheduling strategies evaluated later, where \ac{MCS} adaptation and resource allocation are explicitly modeled.

Packet size variability arises from two main factors:
(i) perception dynamics, which determine the number of included objects; and
(ii) periodic security overhead, due to the periodic certificate transmission.

For $T_{\mathrm{gen}} = 100$~ms, the certificate appears once every ten packets, producing sparse size spikes. For $T_{\mathrm{gen}} = 500$~ms, every second packet carries a certificate, leading to a pronounced alternation between two packet size levels. In addition, variability is more pronounced in the congested scenario due to the higher number of perceived objects.

These observations show that \ac{CPM} traffic exhibits structured and temporally correlated packet size variations, deviating from the quasi-constant traffic assumptions typically underlying semi-persistent scheduling.

\section{NR-V2X Mode~2 Operation and Reaction Mechanisms}

\subsection{Mode~2 Distributed Operation}

\ac{NR}-\ac{V2X} Mode~2 enables distributed sidelink communication without network assistance. All \acp{UE} operate over a shared time--frequency resource pool and autonomously select transmission resources based on channel monitoring and distributed exclusion procedures. By decoding sidelink control information conveying future reservations, \acp{UE} keep track of resources that have been recently used or explicitly reserved by neighboring nodes and exclude them from subsequent selections. This mechanism improves spatial reuse while maintaining predictable timing behavior.

Once selected, a resource may be used for a single transmission or reserved periodically according to a configured \ac{RRI}, forming a semi-persistent allocation. Under \ac{SPS}, the same time--frequency resource is reused across multiple transmission periods. To avoid persistent conflicts, reselection is periodically triggered according to probabilistic rules and counters, resulting in long phases of allocation stability interspersed with occasional reselection events. By limiting the frequency of simultaneous resource selection operations across users, \ac{SPS} enhances distributed allocation predictability and reduces collision exposure.

\subsection{Transport Block Size and Resource Coupling}

A fundamental aspect of Mode~2 operation is the coupling between reserved radio resources and the achievable \ac{TB} size at the MAC layer. In \ac{NR}-\ac{V2X}, resources are allocated on a subchannel basis. For a given transmission, the maximum TB size is determined by the number of allocated subchannels together with the selected \ac{MCS}.

Under \ac{SPS}, a fixed number of subchannels is reserved and reused across consecutive transmissions. For a given \ac{MCS}, this reservation defines a maximum supported \ac{TB} size. \ac{SPS} therefore implicitly assumes that future packets will remain within the \ac{TB} range supported by the reserved subchannel allocation.

When packet size increases beyond the supported \ac{TB} size, the allocation becomes insufficient. Conversely, when packet size decreases, the reserved subchannels may become larger than necessary. In both cases, packet size variability creates a mismatch between the required payload and the capacity supported by the reserved subchannels, directly affecting spectral efficiency and allocation stability.



\subsection{Reaction Mechanisms to Packet Size Variations}

When packet size varies relative to the reserved allocation, the scheduler must react. The available reactions depend on whether packet size increases beyond the supported \ac{TB} size or decreases below it.

\smallskip
\subsubsection{Packet Size Increase}

If the required \ac{TB} size exceeds the maximum size supported by the reserved subchannels and current \ac{MCS}, the \ac{UE} must adapt its transmission. Two principal options are available:

\begin{itemize}
    \item \textbf{Resource reselection:} select a new time--frequency allocation providing additional subchannels.
    \item \textbf{MCS increase (bounded):} increase spectral efficiency within the same subchannel allocation, provided that channel conditions support the higher coding rate.
\end{itemize}

In practice, \ac{MCS} increase is typically bounded by reliability constraints, and a maximum \ac{MCS} is configured to guarantee a minimum target block error rate. As a result, the ability to absorb packet size increases through \ac{MCS} adjustment is limited, and reselection often remains the dominant reaction.

\smallskip
\subsubsection{Packet Size Decrease}

If the packet requires fewer resources than those reserved, additional flexibility exists:

\begin{itemize}
    \item \textbf{Resource reselection:} reduce the allocation by selecting fewer subchannels.
    \item \textbf{Padding:} retain the existing allocation and fill unused resource elements with padding bits.
    \item \textbf{MCS decrease:} lower spectral efficiency within the same subchannel allocation, increasing redundancy and link robustness.
\end{itemize}

These reaction mechanisms are not strictly dictated by the standard and represent implementation choices available to system designers. The selection among them determines the trade-off between resource efficiency, allocation stability, and reliability.
The main reaction mechanisms considered in this work are schematically illustrated in Fig.~\ref{fig:sps_strategies}. The figure focuses on representative cases of resource reselection, padding, and \ac{MCS} optimization under semi-persistent scheduling.

\setlength{\fboxrule}{0.4pt}
\setlength{\fboxsep}{6pt}
\captionsetup[subfloat]{justification=centering,singlelinecheck=false,width=\linewidth}

\begin{figure}[t]
\centering

\noindent\framebox[\columnwidth]{%
  \begin{minipage}[c]{\dimexpr\columnwidth-2\fboxsep-2\fboxrule\relax}
    \centering

    \vspace{0.1cm}

    \subfloat[Reselection to allocate a packet requiring additional subchannels.]{
      \includegraphics[width=0.86\linewidth]{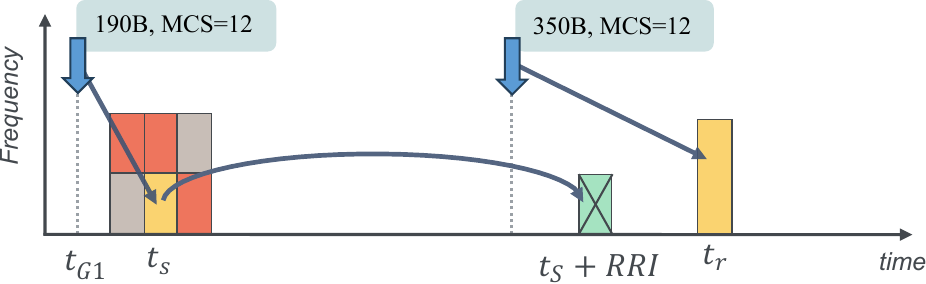}
    }

    \vspace{0.2cm}

    \subfloat[Reselection to reduce allocated subchannels for a smaller packet.]{
      \includegraphics[width=0.86\linewidth]{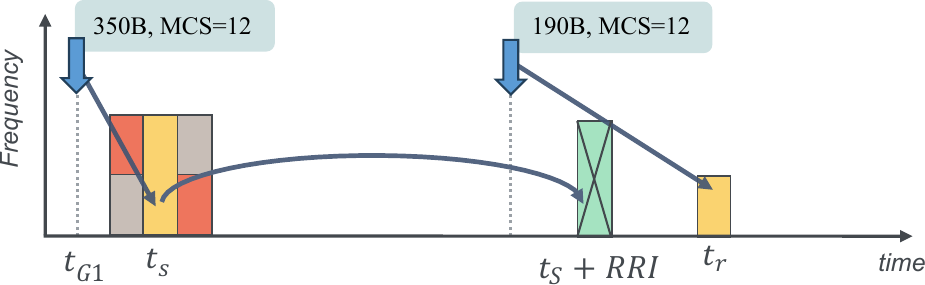}
    }

    \vspace{0.2cm}

    \subfloat[Padding while retaining an oversized reservation.]{
      \includegraphics[width=0.86\linewidth]{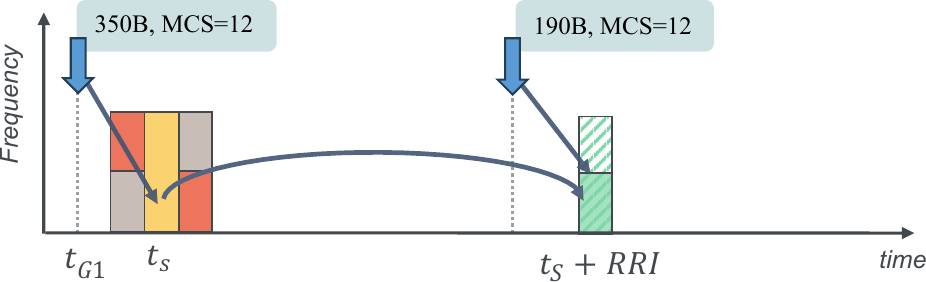}
    }

    \vspace{0.2cm}

    \subfloat[MCS decrease within a fixed subchannel reservation.]{
      \includegraphics[width=0.86\linewidth]{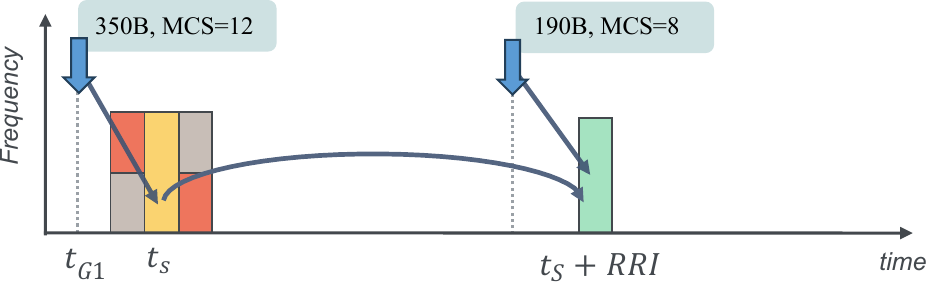}
    }

  \end{minipage}%
}

\vspace{0.2cm}

\subfloat{
  \includegraphics[width=0.86\columnwidth]{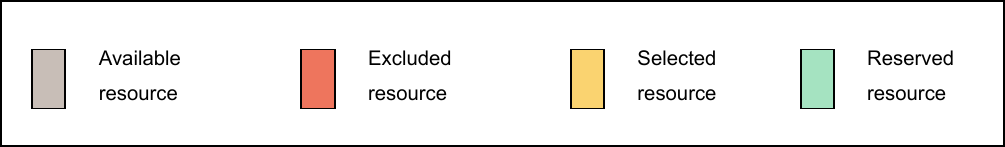}
}

\caption{Reaction mechanisms to packet size variations under semi-persistent scheduling in NR-V2X Mode~2.}
\label{fig:sps_strategies}
\end{figure}

\section{Scheduling strategies considered}\label{Sec:schedulingStrategies}

Based on the reaction mechanisms described in the previous section, we define the scheduling strategies evaluated in this work as specific combinations of responses to packet size variations.

\subsection{Fully Dynamic Scheduling}

In fully \ac{DS}, no semi-persistent reservation is maintained. Resources are selected independently at each transmission, inherently performing reselection for every packet regardless of its size. While this approach maximizes flexibility and naturally accommodates variable packet sizes, it increases selection frequency and reduces allocation predictability.

\subsection{SPS with Aggressive Reselection}

This strategy employs \ac{SPS} with an \ac{RRI}-based reservation but enforces reselection whenever packet size deviates from the reserved allocation, either for increases or decreases. It prioritizes instantaneous resource matching at the expense of allocation stability.

\subsection{SPS with Padding-Based Allocation}

In this approach, reselection is triggered only when packet size exceeds the reserved capacity. When packet size decreases, the allocation is maintained and unused capacity is filled through padding. This strategy prioritizes allocation stability over spectral efficiency.

\subsection{SPS with MCS Adaptation}

This strategy maintains the semi-persistent allocation whenever possible and selects the \ac{MCS} on a per-packet basis within the reserved subchannels, subject to an upper bound given by a configured default (maximum) \ac{MCS}. After each packet generation, the scheduler starts from the default \ac{MCS} and reduces it to the lowest \ac{MCS} that can accommodate the packet within the reserved subchannels. As a result, the selected \ac{MCS} satisfies $m \le m_{\mathrm{def}}$ for all packets, but may fluctuate across packets depending on packet size. Reselection occurs only when the required \ac{TB} size cannot be supported within the reserved subchannels even at $m_{\mathrm{def}}$.

\smallskip
These strategies represent different trade-offs between allocation flexibility, resource stability, and spectral efficiency under distributed \ac{NR}-\ac{V2X} Mode~2 operation. Their impact under realistic variable-size \ac{CPM} traffic conditions is evaluated in the following sections.

\section{System-Level Evaluation Setup}

\subsection{Simulation Framework and Settings}

The scheduling strategies described in Section~IV are evaluated using the open-source WiLabV2XSim framework~\cite{bib:WiLabV2XSim}, extended to support distributed NR-V2X Mode~2 sidelink operation with variable subchannel-based resource allocation, enabling the number of allocated subchannels to adapt to packet size. All UEs autonomously perform resource selection within a shared sidelink bandwidth according to the sensing and exclusion mechanisms described in Section~III. No retransmissions are considered.

The transmission power is $23$~dBm with a $3$~dBi antenna gain and a $9$~dB noise figure. A $20$~MHz channel with $30$~kHz subcarrier spacing is partitioned into 5 subchannels of 10 resource blocks each, with a slot duration of $0.5$~ms. The baseline MCS index is set to 12 (16-QAM, code rate $\approx 0.43$), yielding a maximum transport block size of $1180$ bytes. The channel model follows the rural highway model defined in ETSI TR~103~439. Semi-persistent strategies operate with an RRI equal to the packet generation interval ($T_{\mathrm{gen}}$) and a keep probability $P_{\text{keep}} = 0.8$, while fully dynamic scheduling performs resource selection at every transmission.


\subsection{Road Scenario and Mobility Model}

Simulations are conducted over an 8\,km six-lane highway (three lanes per driving direction). Vehicles are uniformly distributed across lanes and move at constant speed.

Two traffic conditions are considered, corresponding to the non-congested and congested portions of the dataset used for CPM reconstruction:

\begin{itemize}
    \item \textbf{Non-congested highway:} $50$ vehicles/km with an average speed of $110$~km/h and standard deviation $11$~km/h.
    \item \textbf{Congested highway:} $150$ vehicles/km with an average speed of $70$~km/h and standard deviation $7$~km/h.
\end{itemize}

In the simulated highway segment, these densities correspond to 400 and 1200 simultaneously active vehicles, respectively.

\subsection{CPM Traffic Generation}

All vehicles periodically generate \acp{CPM} according to the reconstruction methodology described in Section~II. In the non-congested scenario, the packet generation interval is set to $T_{\mathrm{gen}} = 100$~ms, whereas in the congested scenario $T_{\mathrm{gen}} = 500$~ms is adopted to limit channel load under higher vehicle density.
 The packet sizes reflect both object-driven payload evolution and periodic security-related overhead.

Since a single per-scenario perception trace is available for each traffic condition, the same temporal sequence of detected objects is used across vehicles. To avoid artificial synchronization and preserve temporal correlation properties, each vehicle is assigned a random starting point within the trace. Upon reaching the end of the trace, the sequence is replayed in reverse order, alternating forward and backward playback for the duration of the simulation. This approach maintains temporal structure while introducing inter-vehicle variability in packet sizes.



\subsection{Performance Metrics}

The performance of each scheduling strategy is evaluated using the following metrics:

\begin{itemize}
    \item \textbf{\Ac{PRR}:} the average ratio between the number of vehicles correctly decoding a packet at a given distance from the transmitter and the overall number of vehicles at the same distance;
    \item \textbf{\Ac{CBR}:} the fraction of subchannels in the resource pool for which the sidelink-received signal strength indicator (S-RSSI) measured by the vehicles exceeds a threshold of $-88$ dBm;
    \item \textbf{Reassignment rate:} average number of resource reselection events per vehicle per unit time, used to quantify allocation stability.
\end{itemize}


\section{Performance Evaluation}

We now compare the scheduling strategies defined in Sec.~\ref{Sec:schedulingStrategies} under the realistic variable-size \ac{CPM} traffic model described in Section~II and the simulation setup of Section~V. The results highlight how packet size variability interacts with distributed resource allocation in \ac{NR}-\ac{V2X} Mode~2.

For reference, we also evaluate \ac{SPS} with fixed packet size, where the transport block size is set to the average CPM size observed in each scenario. This configuration represents the idealized case of periodic traffic with constant packet size, matching the design assumptions typically associated with semi-persistent scheduling.

\begin{figure}[t]
  \centering
  \includegraphics[width=0.975\columnwidth]{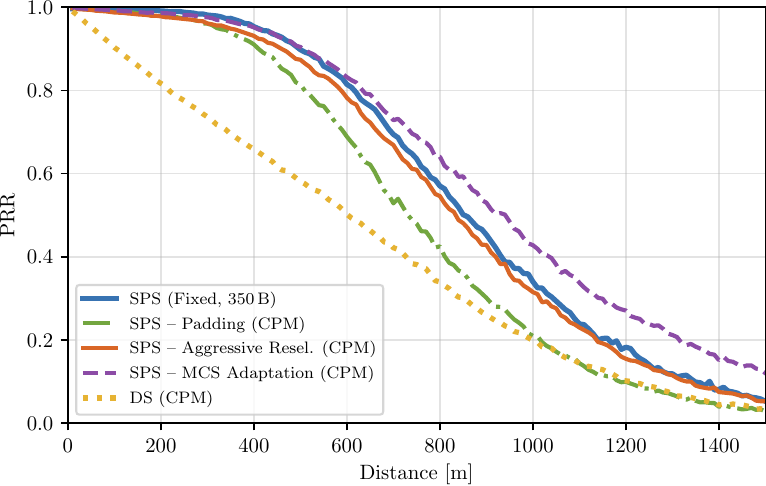}
    \caption{PRR as a function of distance in the non-congested scenario}
  \label{fig:prr_rho50_tgen100}
\end{figure}

\begin{figure}[t]
  \centering
  \includegraphics[width=0.975\columnwidth]{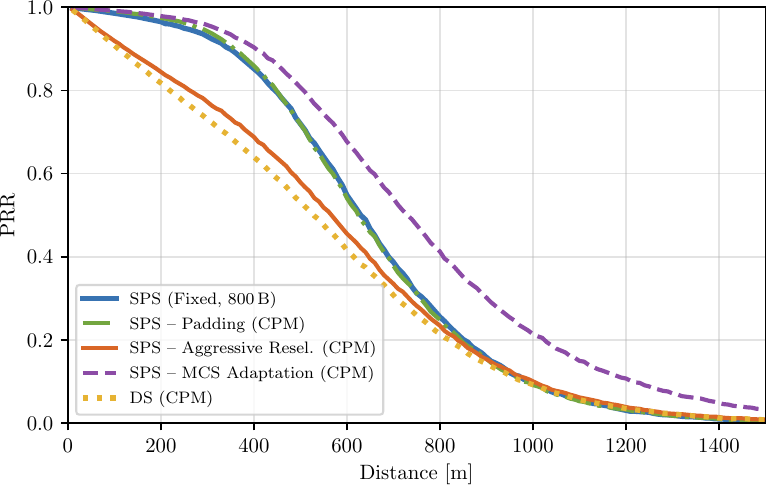}
  \caption{PRR as a function of distance in the congested scenario}
  \label{fig:prr_rho150_tgen500}
\end{figure}

\input{table_metrics.tex}

Table~\ref{tab:prr_metrics} summarizes the average packet size, \ac{CBR}, average number of allocated subchannels, and reassignment rates under both traffic conditions.

\subsection{Impact on Communication Reliability}

Fig.~\ref{fig:prr_rho50_tgen100} and Fig.~\ref{fig:prr_rho150_tgen500} show the \ac{PRR} as a function of communication distance for the different scheduling strategies.

Fully \ac{DS} consistently exhibits the lowest reliability across both traffic conditions. Since resources are reselected for every transmission, allocation stability is minimal and collision probability increases, particularly at medium communication ranges.

\ac{SPS} with aggressive reselection improves instantaneous resource matching by adapting the allocation whenever packet size deviates from the reserved footprint. In the non-congested scenario, this approach achieves slightly better performance than \ac{DS} due to improved resource utilization. However, when traffic density increases, the higher reassignment rate (Table~\ref{tab:prr_metrics}) leads to significant allocation churn, and reliability degrades to levels comparable to \ac{DS}. This confirms that aggressive resource reallocation becomes detrimental in dense scenarios.

The padding-based \ac{SPS} strategy achieves more stable performance across both traffic conditions. By maintaining previously reserved resources whenever possible, it significantly reduces reassignment frequency and limits collision exposure. Despite potential resource underutilization, the improved allocation stability translates into higher \ac{PRR}.

\ac{SPS} with \ac{MCS} adaptation maintains semi-persistent reservations while adjusting spectral efficiency within the reserved subchannels. This strategy avoids unnecessary reselection and achieves reliability close to the padding-based approach, with slight improvements in certain operating points.

\subsection{Allocation Stability and Channel Load}

The reassignment rate reported in Table~\ref{tab:prr_metrics} clearly reflects the stability differences among strategies. \ac{DS} performs reselection at every transmission, while aggressive reselection shows a reassignment rate proportional to packet size fluctuations. Padding and \ac{MCS} adaptation significantly reduce reassignment frequency, preserving allocation continuity.

The measured \ac{CBR} remains broadly comparable across strategies. \ac{DS} and aggressive reselection exhibit slightly lower \ac{CBR} values, particularly in the congested scenario. However, this reduction does not correspond to improved reliability. Instead, it is primarily caused by increased transmission overlaps and collisions, which prevent successful resource usage. Therefore, lower measured \ac{CBR} in these cases reflects inefficient channel usage rather than better spatial reuse.

These observations reinforce that allocation stability, rather than instantaneous resource matching, plays a dominant role in distributed performance.

\subsection{Impact of MCS Adaptation}

\ac{SPS} with \ac{MCS} adaptation adjusts spectral efficiency within the reserved subchannels while preserving the semi-persistent allocation. Compared to aggressive reselection, it maintains low reassignment rates and achieves the highest average \ac{PRR} among the evaluated strategies, particularly in dense conditions.
The performance improvement primarily stems from the increased redundancy applied to smaller packets when a lower MCS is selected. These packets benefit from enhanced robustness without requiring additional subchannels. However, larger packets, which typically determine the reserved allocation size, are often transmitted at the default MCS and do not benefit from MCS adaptation.
Therefore, the reliability gains are not uniform across all packet instances. Since larger packets frequently carry security certificates, their successful reception is essential for validating subsequent messages. Even if smaller packets are received at longer distances due to stronger coding, they may be discarded if the corresponding certificate was not successfully decoded. Consequently, the practical benefit of MCS adaptation may be partially constrained by application-layer dependencies.


\subsection{Discussion}

Across all evaluated scenarios, the results consistently show that \ac{DS} performs worst due to continuous resource reselection. Aggressive reselection can offer moderate gains in low-density scenarios but becomes counterproductive as traffic density increases. In contrast, strategies that minimize reassignment frequency, such as padding-based \ac{SPS} and \ac{MCS} adaptation, achieve more robust performance. By preserving semi-persistent reservations whenever possible, these approaches limit the number of distributed selection events and reduce collision exposure.

Packet size variability alone does not inherently degrade reliability. Rather, the instability induced by frequent resource reallocation is the dominant factor affecting performance. While \ac{MCS} adaptation achieves reliability close to the padding-based strategy and can improve the average \ac{PRR}, its benefits may be limited by the end delivery of larger packets carrying security credentials.

\section{Conclusion}

This paper investigated the impact of realistic variable-size \ac{CPM} traffic on distributed \ac{NR}-\ac{V2X} Mode~2 scheduling. Using perception traces reconstructed from highway driving data, we evaluated fully dynamic scheduling and multiple semi-persistent strategies under identical system conditions. 

The results demonstrate that packet size variability degrades performance primarily when it induces frequent resource reallocations. Conversely, strategies that preserve allocation stability, such as padding-based semi-persistent scheduling, provide more robust reliability in distributed operation. While \ac{MCS} adaptation can improve average performance, its benefits may depend on the interplay between coding robustness, higher-layer validation mechanisms and the end delivery of larger packets
carrying security credentials.

Overall, our findings indicate that maintaining stable semi-persistent resource reservations is generally more beneficial than aggressively optimizing resource matching for each transmission. These insights provide practical guidance for configuring \ac{NR}-\ac{V2X} Mode~2 to support emerging cooperative perception services characterized by structured and temporally correlated traffic variability.

\section*{Acknowledgments}
 This work was supported by the Italian National Recovery and Resilience Plan (NRRP) of NextGenerationEU, partnership on ``Telecommunications of the Future'' (PE00000001 - program ``RESTART''), projects MoVeOver and ALERT.

\bibliographystyle{IEEEtran}
\bibliography{bib_folder/IEEEabrv,bib_folder/bibliography.bib}

\end{document}

%% file: acronyms.tex
\begin{acronym} 
\acro{3GPP}{Third Generation Partnership Project}
\acro{2D}{two-dimensional}
\acro{2G}{second generation}
\acro{3G}{third generation}
\acro{4G}{fourth generation}
\acro{5G}{fifth generation}
\acro{6G}{sixth generation}
\acro{5GAA}{5G Automotive Association}
\acro{5GS}{5G system}
\acro{ACC}{adaptive cruise control}
\acro{AD}{autonomous driving}
\acro{ADAS}{advanced driver assistance systems}
\acro{AF}{application function}
\acro{AGC}{automatic gain control}
\acro{AGV}{automated guided vehicle}
\acro{AI}{artificial intelligence}
\acro{AIFS}{arbitration inter-frame space}
\acro{AMC}{adaptive modulation and coding}
\acro{AoI}{age of information}
\acro{API}{application programming interface}
\acro{AS}{application server}
\acro{AD}{autonomous driving}
\acro{AUTOSAR}{Automotive Open System}
\acro{AWGN}{additive white Gaussian noise}
\acro{BEP}{beacon error probability}
\acro{BM-SC}{broadcast multicast service center}
\acro{BP}{beacon periodicity}
\acro{B-CSA}{broadcast \ac{CSA}}
\acro{BSM}{basic safety message}
\acro{BSS}{basic service set}
\acro{BS}{base station}
\acro{BTP}{Basic Transport Protocol}
\acro{BWP}{bandwidth part}
\acro{C-ACC}{cooperative adaptive cruise control}
\acro{C-GLOSA}{cooperative GLOSA}
\acro{C-ITS}{cooperative-intelligent transport system}
\acro{C-NOMA}{cooperative \ac{NOMA}}
\acro{C-V2X}{cellular-\ac{V2X}}
\acro{C2C-CC}{Car 2 Car Communication Consortium}
\acro{CA}{collision avoidance}
\acro{CACC}{cooperative adaptive cruise control}
\acro{CAM}{cooperative awareness message}
\acro{CAV}{connected and autonomous vehicle}
\acro{CBR}{channel busy ratio}
\acro{CC}{cruise control}
\acro{CCA}{clear channel assessment}
\acro{CCAM}{cooperative, connected and automated mobility}
\acro{CCDF}{complementary cumulative distribution function}
\acro{CDF}{cumulative distribution function}
\acro{CDMA}{code-division multiple access}
\acro{CEPT}{European Conference of Postal and Telecommunications Administrations}
\acro{CoMP}{coordinated multi-point}
\acro{CMM}{co-channel coexistence mitigation methods}
\acro{CN}{core network}
\acro{CP-OFDM}{cyclic prefix orthogonal frequency-division multiplexing}
\acro{CP}{collective perception}
\acro{CPM}{collective perception message}
\acro{CPS}{collective perception service}
\acro{CR}{channel occupancy ratio}
\acro{CRDSA}{contention resolution diversity slotted ALOHA}
\acro{CRLB}{Cram\'er-Rao Lower Bound }
\acro{CSA}{coded-slotted ALOHA}
\acro{CSD}{cyclic shift diversity}
\acro{CSI}{channel state information}
\acro{CSIT}{channel state information at the transmitter}
\acro{CSIR}{channel state information at the receiver}
\acro{CSMA/CA}{carrier sense multiple access with collision avoidance}
\acro{CSMS}{cyber security management system}
\acro{CSSR}{candidate single-subframe resource}
\acro{CTS-To-Self}{Clear-To-Send-To-Self}
\acro{DA}{data age}
\acro{D2D}{device-to-device}
\acro{DC}{duty cycle}
\acro{DCC}{decentralized congestion control}
\acro{DCF}{distributed coordination function}
\acro{DCI}{downlink control
information}
\acro{DCM}{dual carrier modulation}
\acro{DENM}{decentralized environmental notification message}
\acro{DMRS}{demodulation reference signal}
\acro{DoS}{denial-of-service}
\acro{DDoS}{distributed denial-of-service}
\acro{DOT}{Department of Transportation}
\acro{DRX}{discontinuous reception}
\acro{DSRC}{Dedicated Short Range Communication}
\acro{DS}{dynamic scheduling}
\acro{DT}{digital twin}
\acro{DtS}{direct-to-satellite}
\acro{DtS-IoT}{direct-to-satellite Internet of Things}
\acro{EC}{European Commission}
\acro{ECP}{extended cyclic prefix}
\acro{EDCA}{enhanced distributed coordination access}
\acro{EDGE}{Enhanced Data rates for GSM Evolution}
\acro{EED}{eEnd-to-end delay}
\acro{eMBMS}{evolved Multicast Broadcast Multimedia Service}
\acro{eMBB}{enhanced mobile broadband}
\acro{eNodeB}{evolved NodeB}
\acro{EPC}{Evolved Packet Core}
\acro{EPS}{Evolved Packet System}
\acro{ES}{energy signal}
\acro{ETSI}{European Telecommunications Standards Institute}
\acro{eV2X}{enhanced V2X}
\acro{FANET}{flying ad hoc network}
\acro{FCC}{Federal Communications Commission}
\acro{FCD}{floating car data}
\acro{FD}{full-duplex}
\acro{FDD}{frequency division duplex}
\acro{FDMA}{frequency division multiple access}
\acro{FEC}{forward error correction}
\acro{FR1}{frequency range 1}
\acro{FR2}{frequency range 2}
\acro{GapRe}{gap sensing and resource re-selection}
\acro{GLOSA}{green light optimal speed advisory}
\acro{GNSS}{global navigation satellite system}
\acro{GNW}{GeoNetworking}
\acro{GPRS}{general packet radio service }
\acro{GPS}{global positioning system}
\acro{HARQ}{hybrid automatic repeat request}
\acro{HD}{high-definition}
\acro{HPC}{high performance computing}
\acro{HRLLC}{hyperreliable
and low-latency communication}
\acro{HSDPA}{High Speed Downlink Packet Access}
\acro{HSPA}{High Speed Packet Access}
\acro{IBE}{in-band emission}
\acro{ICI}{inter-carrier interference}
\acro{IDMA}{interleave-division multiple access}
\acro{IEEE}{Institute of Electrical and Electronics Engineers}
\acro{IM}{index modulation}
\acro{IoT}{Internet of Things}
\acro{IP}{Internet Protocol}
\acro{IPG}{inter-packet gap}
\acro{IR}{incremental redundancy}
\acro{IRS}{intelligent reflective surface}
\acro{ISAC}{integrated sensing and communication}
\acro{ISI}{inter-symbol interference}
\acro{ISM}{industrial, scientific, and medical}
\acro{ISO}{international standardization organization}
\acro{ITU-R}{International Telecommunication Union Radiocommunication Sector}
\acro{ITS}{intelligent transport system}
\acro{i.i.d.}{independent identically distributed}
\acro{KPI}{key performance indicator}
\acro{LBT}{listen-before-talk}
\acro{LDPC}{low-density parity-check}
\acro{LDS}{low-density spreading}
\acro{LEO}{low Earth orbit}
\acro{LIDAR}{light detection and ranging}
\acro{LLR}{log-likelihood ratio}
\acro{LOD}{Linked Open Data}
\acro{LOS}{line-of-sight}
\acro{LoRa}{Long Range}
\acro{LoRa-CSS}{long range chirp spread spectrum}
\acro{LoRaWAN}{Long Range Wide Area Network}
\acro{LPWAN}{low-power wide-area network}
\acro{LR-FHSS}{long range frequency hopping spread spectrum}
\acro{LTE}{long term evolution}  
\acro{LTE-D2D}{\ac{LTE} with \ac{D2D} communications}  
\acro{LTE-LAA}{LTE license assisted access}
\acro{LTE-V2X}{long-term-evolution-vehicle-to-anything} 
\acro{LTE-V2V}{\ac{LTE}-\ac{V2V}}
\acro{LwM2M}{lightweight machine-to-machine} 
\acro{MAC}{medium access control}
\acro{MANET}{mobile ad hoc network}
\acro{MBMS}{Multicast Broadcast Multimedia Service}
\acro{MBMS-GW}{MBMS Gateway}
\acro{MBSFN}{MBMS Single Frequency Network}
\acro{MCE}{Multi-cell Coordination Entity}
\acro{MEC}{mobile edge computing}
\acro{MCM}{maneuver coordination message}
\acro{MCS}{modulation and coding scheme}
\acro{MIMO}{multiple input multiple output}
\acro{ML}{machine learning}
\acro{mMTC}{massive machine-type communications}
\acro{MRC}{maximum ratio combining}
\acro{MRD}{maximum reuse distance}
\acro{MUD}{multi-user detection}
\acro{NAV}{network allocation vector}
\acro{NB-IoT}{Narrowband IoT}
\acro{NEF}{network exposure function}
\acro{NGSI-LD}{next generation service interfaces - linked data}
\acro{NGV}{Next Generation V2X}
\acro{NHTSA}{National Highway Traffic Safety Administration}
\acro{NIS2}{network and information systems directive}
\acro{NLOS}{non-line-of-sight}
\acro{NOMA}{non-orthogonal multiple access}
\acro{NOMA-MCD}{\ac{NOMA}-mixed centralized/distributed}
\acro{NR}{New Radio}
\acro{NR-U}{NR unlicensed}
\acro{NTN}{non-terrestrial network}
\acro{OBU}{on-board unit}
\acro{OCB}{outside of the context of a basic service set}
\acro{OEM}{original equipment manufacturer}
\acro{OFDM}{orthogonal frequency-division multiplexing}
\acro{OFDMA}{orthogonal frequency-division multiple access}
\acro{OMA}{orthogonal multiple access}
\acro{OTFS}{orthogonal time frequency space}
\acro{pdf}{probability density function}
\acro{PAPR}{peak to average power ratio}
\acro{PCF}{policy control function}
\acro{PCM}{platoon control message}
\acro{PD}{packet delay}
\acro{PEP}{pairwise error probability} 
\acro{PER}{packet error rate}
\acro{PIAT}{packet inter-arrival time}
\acro{PLS}{physical layer security}
\acro{PPDU}{physical layer protocol data unit}
\acro{PRB}{physical resource block}
\acro{PSCCH}{Physical Sidelink Control Channel}
\acro{PSDU}{physical layer convergence protocolservice data unit}
\acro{PSFCH}{Physical Sidelink Feedback channel}
\acro{PSSCH}{Physical Sidelink Shared channel}
\acro{PHY}{physical}
\acro{PL}{path loss}
\acro{PLMN}{Public Land Mobile Network}
\acro{PPPP}{ProSe Per-Packet Priority}
\acro{PPPR}{ProSe Per-Packet Reliability}
\acro{ProSe}{Proximity-based Services}
\acro{PRR}{packet reception ratio}
\acro{QAM}{quadrature amplitude modulation}
\acro{QC-LDPC}{quasi-cyclic \ac{LDPC}}
\acro{QoS}{quality of service}
\acro{QPSK}{quadrature phase shift keying}
\acro{RAN}{radio access network}
\acro{RAT}{radio access technology}
\acro{RB}{resource block}
\acro{RCS}{radar cross-section}
\acro{RE}{resource element}
\acro{RF}{radio frequency}
\acro{RIS}{reconfigurable intelligent  surfaces}
\acro{RL}{reinforcement learning}
\acro{RNN}{recurrent neural network}
\acro{RR}{radio resource}
\acro{RRC}{radio resource control}
\acro{RRI}{resource reservation interval}
\acro{RSRP}{reference signal received power}
\acro{RSSI}{Received Signal Strength Indicator}
\acro{RSU}{road side unit} 
\acro{SAE}{Society of Automotive Engineers}
\acro{SAI}{Service Area Identifier}
\acro{SB-DS}{sensing-based dynamic scheduling}
\acro{SB-SPS}{sensing-based semi-persistent scheduling}
\acro{SC-FDMA}{single carrier frequency division multiple access}
\acro{SC-PTM}{Single Cell Point To Multipoint}
\acro{SCS}{subcarrier spacing}
\acro{SCMA}{sparse-code multiple access}
\acro{SI}{self-interference}
\acro{SIC}{successive interference cancellation}
\acro{SIFS}{short inter-frame space}
\acro{S-UE}{scheduling UE}
\acro{SCI}{sidelink control information}
\acro{SDG}{sustainable development goals}
\acro{SDN}{software-defined networking}
\acro{SF}{spreading factor}
\acro{SFFT}{symplectic finite Fourier transform}
\acro{SHINE}{simulation platform for heterogeneous interworking networks}
\acro{SINR}{signal-to-interference-plus-noise ratio}
\acro{SNR}{signal-to-noise ratio}
\acro{SRS}{sounding reference signal}
\acro{SPS}{semi-persistent scheduling}
\acro{STBC}{space time block codes}
\acro{SV}{smart vehicle}
\acro{TB}{transport block}
\acro{TBC}{time before change}
\acro{TBE}{time before evaluation}
\acro{TDD}{time-division duplex}
\acro{TDMA}{time-division multiple access}
\acro{TR}{transmission range}
\acro{TS}{Technical Specification}
\acro{TTI}{Transmission Time Interval}
\acro{UAV}{uncrewed aerial vehicle}
\acro{UE}{User Equipment}
\acro{UD}{update delay}
\acro{UDD}{urban digital development}
\acro{UMTS}{universal mobile telecommunications system}
\acro{URI}{Uniform Resource Identifier} 
\acro{URLLC}{ultra-reliable and ultra-low latency communications}
\acro{USIM}{Universal Subscriber Identity Module}
\acro{UTDOA}{uplink time difference of arrival}
\acro{UWAC}{underwater acoustic communications}
\acro{V2C}{vehicle-to-cellular} 
\acro{V2N}{vehicle-to-network}
\acro{V2I}{vehicle-to-infrastructure} 
\acro{V2P}{vehicle-to-pedestrian}
\acro{V2R}{vehicle-to-roadside}
\acro{V2V}{vehicle-to-vehicle} 
\acro{V2X}{vehicle-to-everything} 
\acro{VoI}{value of information}
\acro{VANET}{vehicular ad hoc network}
\acro{VAM}{VRU awareness message}
\acro{VRU}{vulnerable road user}
\acro{WAVE}{wireless access in vehicular environment}
\acro{WBSP}{wireless blind spot probability}
\acro{XML}{extensible markup language}
\acro{SL}{sidelink}
\acro{ECC}{Electronic Communications Committee}
\end{acronym}

%% file: table_metrics.tex
\begin{table}[t]
\caption{Performance metrics under non-congested and congested scenarios.}
\label{tab:prr_metrics}
\centering
\footnotesize
\setlength{\tabcolsep}{3.2pt}
\renewcommand{\arraystretch}{1.15}
\begin{tabular}{lccccc}
\hline
\hline
\rule{0pt}{5ex}\shortstack[t]{Policy\\{}} &
\shortstack[t]{Avg Pkt.\\Size [B]} &
\shortstack[t]{CBR\\{[\%]}} &
\shortstack[t]{Avg\\Subch.} &
\shortstack[t]{Reass.\\{/s}} &
\shortstack[t]{Size\\Reass./s} \\
\hline
\multicolumn{6}{c}{\textit{Non-congested scenario}} \\
\hline
SPS (Fixed, 350\,B)        & 350 & 21 & 2.00 & 0.24 & 0.00 \\
SPS -- Padding             & 308 & 24 & 2.60 & 0.34 & 0.12 \\
SPS -- Aggressive Resel.   & 308 & 21 & 2.09 & 1.27 & 1.11 \\
SPS -- Adapt MCS           & 308 & 24 & 2.60 & 0.34 & 0.12 \\
DS                         & 308 & 20 & 2.09 & 9.92 & 0.00 \\
\hline
\multicolumn{6}{c}{\textit{Congested scenario}} \\
\hline
SPS (Fixed 800\,B)         & 800 & 18 & 4.00 & 0.06 & 0.00 \\
SPS -- Padding             & 708 & 19 & 4.19 & 0.09 & 0.04 \\
SPS -- Aggressive Resel.   & 708 & 16 & 3.63 & 1.38 & 1.36 \\
SPS -- Adapt MCS           & 708 & 19 & 4.19 & 0.09 & 0.04 \\
DS                         & 708 & 16 & 3.63 & 1.99 & 0.00 \\
\hline
\hline
\end{tabular}
\end{table}